\documentclass[fleqn,usenatbib]{mnras_mod}

\usepackage{txfonts}

\usepackage[T1]{fontenc}

\usepackage{graphicx}    
\usepackage{amssymb}    
\usepackage{enumerate}
\usepackage{hyperref}
\usepackage{lineno}

\newcommand{\aref}[1]{\hyperref[#1]{Appendix~\ref{#1}}}

\newcommand\blfootnote[1]{%
  \begingroup
  \renewcommand\thefootnote{}\footnote{#1}%
  \addtocounter{footnote}{-1}%
  \endgroup
}

\renewcommand{\footnoterule}{%
  \kern -3pt
  \hrule width 0.2\textwidth height 1pt
  \kern 2pt
}

\hypersetup{linkcolor=red}          
\title[]{{\Large Article}\\ \vspace{-15pt} \line(1,0){500}\\Prevalent elongated galaxies in the early Universe evidenced by stellar kinematics}

\author[B. Wang et al.]{
	Bitao Wang,$^{1}$
	Yingjie Peng,$^{2,3}$
  Hua Gao$^{4}$
	\\
}

\date{}

\pubyear{2025}

\begin{document}
\label{firstpage}
\pagerange{\pageref{firstpage}--\pageref{lastpage}}
\maketitle


\textbf{
The Universe is now extensively populated by discy galaxies with coherent galaxy-wise stellar rotation\cite{2008MNRAS.388.1321P, 2011MNRAS.414.2923K, 2013MNRAS.432.1709C, 2020MNRAS.495.1958W}.
This disc prevalence has been deemed a late-time phenomenon because the penetrating cold gaseous streams\cite{2006MNRAS.368....2D} in the early Universe ($z\gtrsim 2$) fuel the star formation in galaxies too intensively to allow for thin disc formation\cite{2024MNRAS.532.3808M}.
However, recent images taken by the James Webb Space Telescope (JWST) unveiled a prominent population of low-mass galaxies at high redshifts with flattened shapes, widely interpreted as early significance of discs\cite{2022ApJ...938L...2F, 2024ApJ...966..113L} given the well-established connection between flattening and discy morphology seen in the local Universe.
It is noticed, on the other hand, that these galaxies show far more flattened systems than can be accounted for by randomly oriented oblate discs\cite{2006ApJ...652..963R, 2012ApJ...745...85L}, and the axial ratio distributions are better explained by elongated prolate ellipsoids\cite{2014ApJ...792L...6V, 2019MNRAS.484.5170Z, 2024ApJ...963...54P, 2024ApJ...973L..29W}, an extremely rare spindle-like configuration at low redshifts.
The true morphological nature of these early low-mass galaxies is fundamental to understanding the structure evolution of their discy descendants we see today, including our Milky Way.
In this work, we discriminate the oblate disc and prolate spindle scenario by a decisive experiment with stellar kinematics at its core.
The result clearly supports the prolate spindle scenario, and evidences an early Universe widely inhabited by linear stellar systems contrasting the current era dominated by planar discy galaxies, which suggests a dimensional transition in galactic structure over cosmic time.
}

\blfootnote{$^{1}$School of Physics and Electronics, Hunan University, Changsha 410082, China $^{2}$Department of Astronomy, School of Physics, Peking University, Beijing 100871, China $^{3}$Kavli Institute for Astronomy and Astrophysics, Peking University, Beijing 100871, China $^{4}$School of Physics and Astronomy, University of Nottingham, University Park, Nottingham NG7 2RD, UK}

\begin{figure*}
	\includegraphics[width=1.05\textwidth]{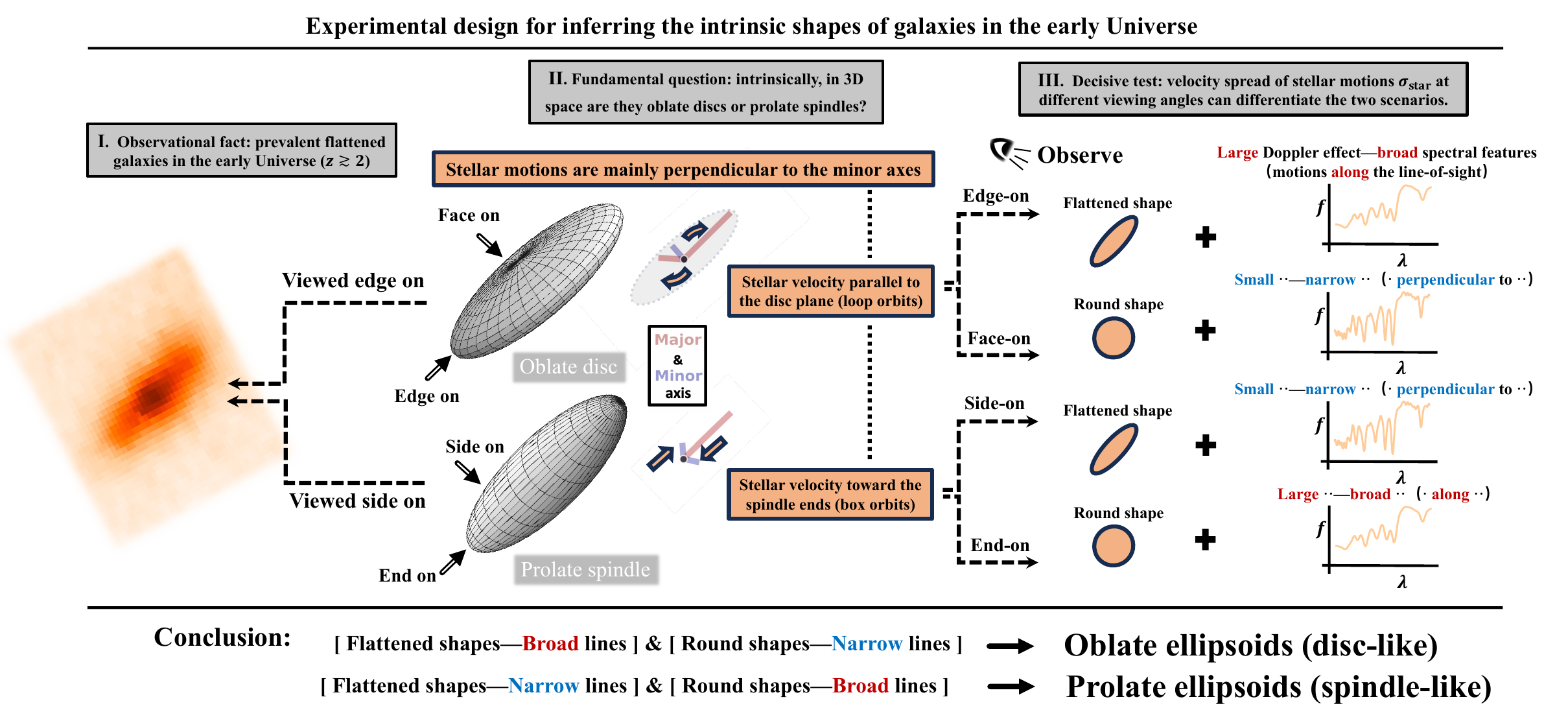}
	\caption{\textbf{A schematic plot for the astrophysical experiment based on stellar kinematics that decisively differentiates the two possible configurations for the intrinsic shapes of the prevalent flattened galaxies at high redshifts revealed by JWST.}
  The reddish cutout on the left shows a typical flattened galaxy seen in JWST images.
  While it could be an oblate disc galaxy viewed relatively edge on, a common case at low redshifts, this galaxy is also likely to be a prolate spindle galaxy with its side toward the observer as illustrated by the 3D figures.
  The light red and blue lines to the right of the 3D figures mark their major and minor axes respectively in three-dimensional space.
  The dramatically different geometries of an oblate disc and a prolate spindle galaxy are determined by the distinct kinematics of their stars which is summarized in the orange boxes.
  A critical prediction of such distinct stellar kinematics and the associated intrinsic shapes is that oblate disc and prolate spindle galaxies will manifest contrary apparent shapes versus observed velocity dispersion relation as illustrated in the part III.
  The observed relation is shown in \autoref{fig:2}.
	}
	\label{fig:1}
\end{figure*}

Observations of stellar kinematics have transformed the views on the intrinsic structure and morphology of local galaxy populations\cite{2016ARA&A..54..597C}.
One of the most significant discoveries is that about two thirds of elliptical galaxies, which were thought to be ellipsoidal in three dimensional space, are actually more like spiral discy galaxies in terms of intrinsic morphology\cite{2011MNRAS.414..888E}.
Dynamically hot ellipsoids are the minority unless galaxies are extremely massive ($\mathcal{M}_{\star}\gtrsim10^{11}\mathcal{M}_{\odot}$) and with low star formation rates\cite{2018MNRAS.477.4711G, 2021MNRAS.505.3078V, 2025NatAs...9..165W}.

The three dimensional morphology of early galaxy populations at high redshifts, especially of the stellar component, remains much more obscure.
Various modern cosmological simulations all suggest that, unlike the current galaxy population, dynamically cold stellar discs are rare among early galaxy populations before the cosmic noon $z\gtrsim 2$\cite{2019MNRAS.483..744T, 2019MNRAS.487.5416T, 2019MNRAS.490.3196P}.
Therefore, the observed significant population of early disc-like galaxies indeed potentially pose a challenge to theories, if these galaxies are confirmed to be discs.
The alternate prolate spindle scenario, to account for the large fraction of flattened galaxies, will be equally striking considering the near absence of highly elongated prolate ellipsoids that are required to fit the JWST data in the local Universe.

Inherently, the oblate disc and prolate spindle morphological configurations correspond to drastically different kinematics of stars.
We take advantage of such a Rosetta Stone, and devise an astrophysical experiment that can offer key empirical evidence for discriminating the two scenarios.
The essence of this experiment is illustrated in \autoref{fig:1}.

It shows on the left an oblate disc, with one minor axis and two major axes that form the disc plane, and a prolate spindle with two minor axes and one major axis. They can both manifest a flattened shape when projected at appropriate viewing angles (i.e. relatively edge-on discs and side-on spindles).
The momenta of stellar populations are mainly perpendicular to the minor axes, corresponding to the length differences between the principal axes.
In oblate discs, stars thus move parallel to the disc plane to the first order, while in prolate spindles, generally the motions are radially toward the spindle ends.
As a result, the stellar motions of a disc galaxy would be largely perceived when it is observed relatively edge on (motions at velocities high and low $\Rightarrow$ large velocity dispersion $\sigma_\mathrm{star}$), and thus in an apparently flattened shape.
When the disc is seen face on, showing a round shape, weak vertical motions perpendicular to the disc plane would then fall aligned with the line of sight (i.e. small velocity values $\Rightarrow$ little $\sigma_\mathrm{star}$).
The case for a spindle galaxy is totally in the contrary.
The majority of its stellar motions can be observed if the spindle is seen end on, in its apparently round shape.
Such critically contrary relations between projected shapes (measured via light spatial distributions) versus observed stellar velocity dispersion (probed by the Doppler broadening of spectral features) for the two configurations, as concluded in the right part of \autoref{fig:1}, can be readily compared with the real situation in JWST imaging and spectroscopic data.

We conduct this experiment based on the Data Release 3 of the JWST Advanced Deep Extragalactic Survey\cite{2023arXiv230602465E, 2025ApJS..277....4D}, which provides both imaging and spectroscopy for the largest statistical sample of galaxies in the early Universe.
We take the fully reduced 1D grating spectra covering 0.6–5.3 $\mu$m at spectral resolution $R\sim 1000$, and select those with redshifts robustly measured via detected strong lines (flagged "A", $1/3$ of the whole sample) to minimize errors in measuring stellar velocity dispersion.
The luminous galaxy populations at high redshifts are rapidly growing and still in their low-mass early stage, with typical stellar masses $\mathrm{lg}(\mathcal{M}_{\star}/\mathcal{M}_{\odot})\sim9$ as in the JWST major samples\cite{2023ApJ...946L..15K, 2023ApJ...955...94F, 2025ApJS..277....4D} of early galaxies in redshift range $z\in (1.5,4)$, which are also the observed dominant population that show excessive flattened systems.
We limit our experiment to a stellar mass\cite{2015ApJ...801...97S, 2019ApJS..243...22B} bin $9<\mathrm{lg}(\mathcal{M}_{\star}/\mathcal{M}_{\odot})<9.5$ in the above redshift range, to exclude the mass dependence of stellar velocity dispersion and to balance between signal-to-noise ratio (SNR) and sample size, reaching a final sample of 122 galaxies at an average redshift $\langle z \rangle=2.8$.

Using a dedicated photometric catalogue\cite{2025A&A...699A.343G} for JWST sources, the sample is divided into apparently round and flattened subsample by the median minor-to-major axial ratio $b/a$ of about 0.5 (inner 68\% of the subsample $b/a$ distribution: round, 0.53-0.75 and flattened, 0.31-0.45).
The depths of current data prohibit reliably resolving stellar absorption lines for a statistical subsample.
Therefore, we utilize the extensively applied and tested stacking technique to build high-SNR composite spectra that represent the general characteristics of subsamples.
We cull the restframe 370-415 nm with abundant strong Balmer and metal lines, and extract the median of normalized (by spectrum median flux level) individual spectra as a robust measure against outliers\cite{2020MNRAS.499.1721C}.
The stellar components of the median spectra, with emission lines masked, are then modeled through the full spectrum-fitting method \textsc{ppxf}\cite{2004PASP..116..138C, 2017MNRAS.466..798C}, using a large library of 559 \textsc{bpass}\cite{2022MNRAS.512.5329B} stellar templates on a 13 metallicity $\times$ 43 age grid---metal mass fractions $Z$ ranging from extremely metal-poor value 0.001\% to supersolar value 4\% and ages logarithmically spaced by 0.1 dex from 1 Myr to 15.85 Gyr.
The templates assume an initial mass function (IMF) with a power-law slope $ - 2.35$ above half solar mass and a slope $ - 1.3$ at lower masses.
A Gaussian line-of-sight velocity distribution of stars is implemented in the models to explain the broadening of spectral features.
We use a flexible four-parameter attenuation function\cite{2023MNRAS.526.3273C}, and 10th-order additive polynomials\cite{2019AJ....158..231W} to account for calibration residuals beyond our measure.
The best fit kinematics of stellar populations, the focus of our experiment, is insensitive to the specific choices for the model ingredients.

\begin{figure*}
	\includegraphics[width=0.8\textwidth]{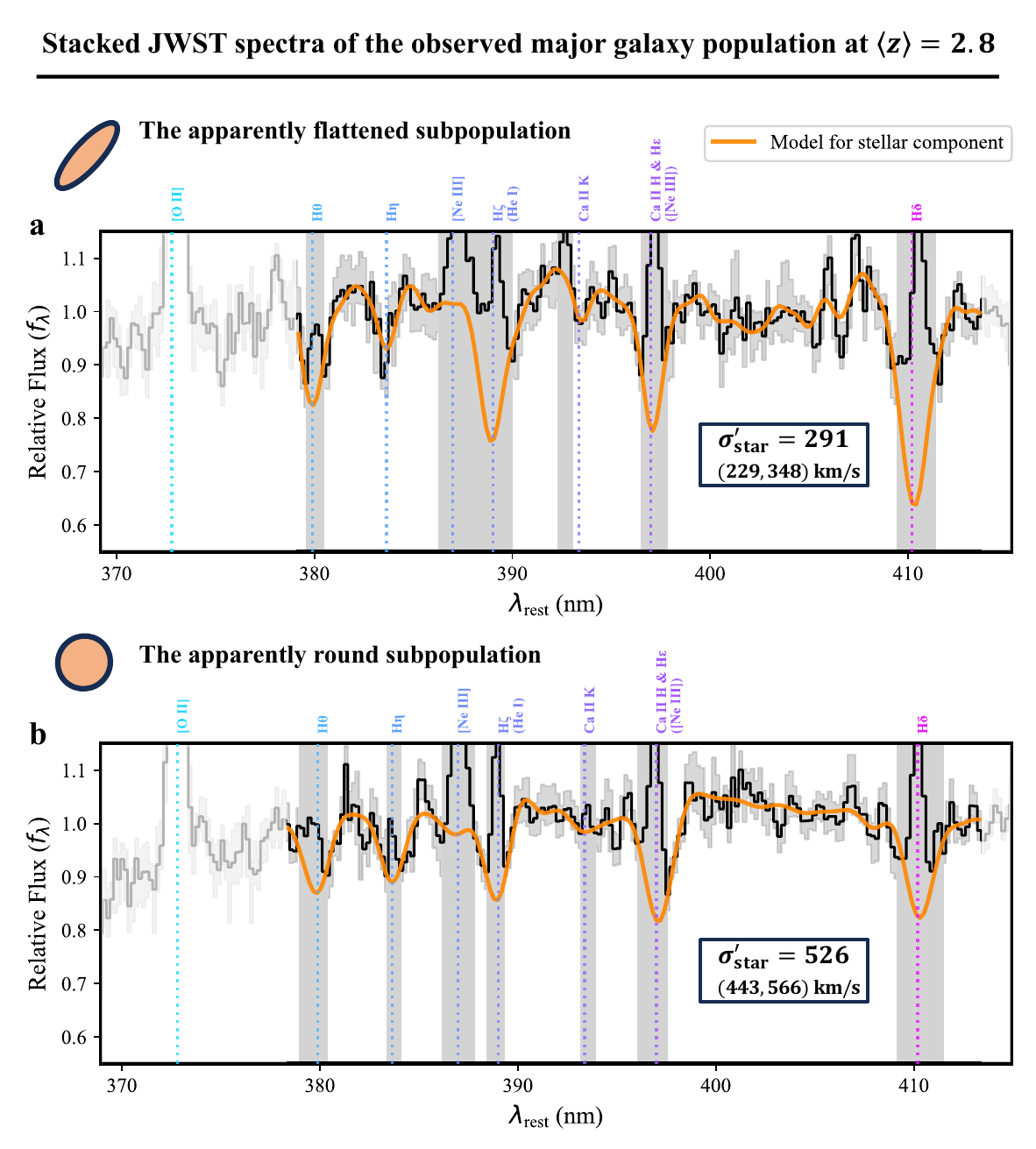}
	\caption{\textbf{The observed apparent shape vs. stellar velocity dispersion relation for the major galaxy population in the JWST high-$z$ sample with stellar masses $\mathrm{lg}(\mathcal{M}_{\star}/\mathcal{M}_{\odot})\sim9$.}
  The experiment is conducted specifically with 122 galaxies in redshift range $z\in (1.5,4)$ and in a narrow stellar mass bin $9<\mathrm{lg}(\mathcal{M}_{\star}/\mathcal{M}_{\odot})<9.5$, which are further divided into apparently flattened (upper panel) and round (lower panel) subpopulation by their median minor-to-major axial ratio.
  In each panel, the black solid line and its associated gray band represent the stacked spectrum and its one-sigma uncertainty estimated via 100 times bootstrap resampling.
  The positions of prominent gaseous emission and stellar absorption lines are marked by vertical dotted lines, where in many cases the line emission and absorption are superimposed.
  The bracketed He I and [Ne III] denote their possible existence underlying the observed emission lines.
  The orange line shows the best fit stellar component in the spectrum, with the emission lines masked, and the measured stellar velocity dispersion $\sigma^{\prime}_\mathrm{star}$ and its 1000 times bootstrapping estimated 68\% confidence interval are recorded in the small inset panel.
  We note that, as compared to the intrinsic values $\sigma_\mathrm{star}$, the measurements $\sigma^{\prime}_\mathrm{star}$ are systematically enlarged by the line broadening due to finite spectral resolution and statistical uncertainties in the measured redshifts for individual galaxies, both of which do not explicitly depend on the apparent shapes of galaxies (more discussion in the main text).
	}
	\label{fig:2}
\end{figure*}

The observations and the model fits for the flattened and round subsample are shown respectively in the upper and lower panel of \autoref{fig:2}.
In each panel, the black solid line and its associated gray band represent the median of the spectra stack and its one-sigma uncertainty estimated via bootstrap resampling.
The positions of prominent spectral features, including both gaseous emission and stellar absorption, are marked by vertical dotted lines, where in many cases the line emission and absorption are superimposed.
The bracketed He I and [Ne III] denote their possible contribution to the observed emission lines.
The modeled stellar component (orange line) matches the observation within uncertainties, and accurately captures the wings of stellar absorption lines that are not impacted by line emission.
The measured stellar velocity dispersions $\sigma^{\prime}_\mathrm{star}$ and their 68\% confidence intervals, estimated based on 1000 wild bootstrap\cite{DAVIDSON2008162} samples, are recorded in the inset panels showing that $\sigma^{\prime}_\mathrm{star}$ of the apparently round subpopulation is considerably larger than that of the flattened subpopulation.

To assess the influence of masking the infilling emission lines over measuring stellar velocity dispersion, we run a Monte Carlo simulation of ten thousand mock spectra.
Each spectrum in the same wavelength range as the analyzed JWST spectra is synthesized with 10\% \textsc{bpass} stellar templates randomly selected each time, sampling their weights uniformly in (0,1), and with the five prominent Balmer emission lines modelled by Gaussians whose equivalent widths range from (-0.4, -1.2) times the equivalent widths of the superimposing absorption lines.
The stellar templates are convolved by a Gaussian velocity distribution of dispersion $\sigma_\mathrm{star}$ uniformly sampled from 200-600 km/s while emission lines by $\sigma_\mathrm{gas}$ randomly picked in the range (0.5$\sigma_\mathrm{star}$, 0.9$\sigma_\mathrm{star}$) given the support of gas, in addition to by motions, by pressure due to magnetic field and cosmic rays (Pipe3D measurements\cite{2022ApJS..262...36S} show $\sigma_\mathrm{gas}\sim 0.85 \sigma_\mathrm{star}$ for the central region of 2.5" diameter among general galaxy populations at redshift zero observed by MaNGA survey\cite{2015ApJ...798....7B, 2022ApJS..259...35A}.).
Extended Data Fig.1 shows that, by masking the infilling emission lines and fitting the clearly observable absorption wings, precise measurements of stellar dispersion can still be achieved with typical errors at the level of several percent.

Besides masking the emission lines, we also tried fitting the infilling emission lines (Gaussian profiles) and the stellar absorptions simultaneously.
The uncertainties are larger as compared to fitting the absorption wings alone but results are still consistent within bootstrapped errors: the measured velocity dispersions $\sigma^{\prime}_\mathrm{star}$ in this case are 334 (265, 373) km/s for the apparently flattened subsample and 449 (412, 576) km/s for the apparently round subsample.

\begin{figure*}
	\includegraphics[width=0.7\textwidth]{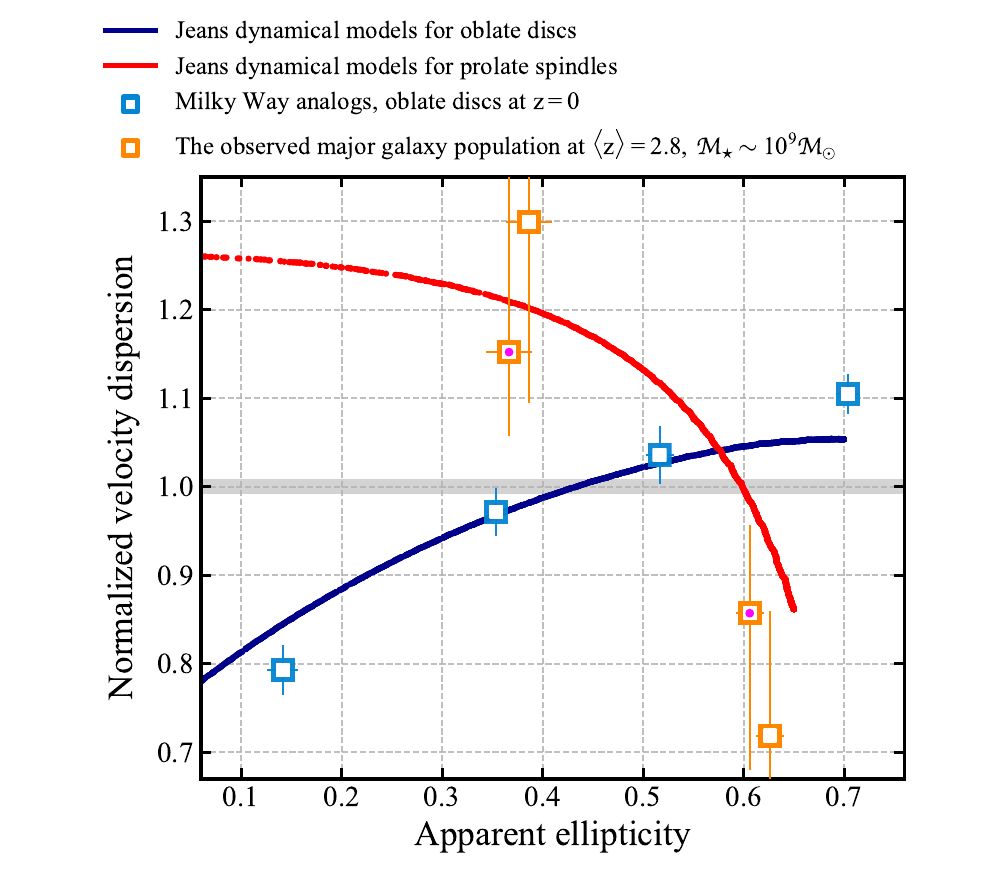}
	\caption{\textbf{The observations and theoretical predictions for the apparent shape vs. stellar velocity dispersion relations of oblate discs and prolate spindles.}
  The red and blue tracks show how the effective stellar velocity dispersion $\sigma_\mathrm{e}$ varies with apparent ellipticity (2000 randomly sampled inclination angles) predicted by Jeans Anisotropic Models respectively for prolate and oblate ellipsoids of exponential light profiles, normalized to the $\sigma_\mathrm{e}$ at their median ellipticity.
  The shapes of these two tracks are not sensitive to the anisotropy level of dynamical models, and the clear discontinuity of the red track at low ellipticities reflect the fact that randomly oriented prolate ellipsoids are highly skewed toward flattened shapes.
  The blue squares and the error bars mark the median normalized $\sigma_\mathrm{e}$ and ellipticities with their 1000 times bootstrapping estimated 68\% confidence intervals in the four quarters of the ellipticity distribution for the Milky Way analogs in MaNGA survey.
  The orange squares stand for the apparently round (the left two) and flattened (the right two) subsample of the analyzed JWST galaxies, with emission lines included in the fit (marked by central magenta dot) or masked.
  The measured velocity dispersion $\sigma^{\prime}_\mathrm{star}$ and uncertainties are normalized to the linearly interpolated value at the median ellipticity 0.5 of the whole sample, for the purpose of qualitatively illustrating the trend.
	}
	\label{fig:3}
\end{figure*}

Two significant effects exist and make the measured velocity dispersion systematically larger than the intrinsic velocity dispersion of stars $\sigma_\mathrm{star}$.
The first is the instrument broadening of spectral lines due to the finite wavelength resolution ($R=\lambda/\Delta\lambda \sim 1000$) of JWST grating spectra. This effect results in a velocity spread of $\sigma_\mathrm{inst}\sim 127\ \mathrm{km/s}$ in terms of standard deviation.
The second originates from the errors in the measured redshifts of individual galaxies.
In spectra stacking, features of the composite spectrum are broadened as the errors aggregate, and this source of line broadening is estimated to be at the order of $100\ \mathrm{km/s}$ considering the wavelength calibration error and the uncertainties in the redshift determination processes\cite{2025ApJS..277....4D}.
An accurate correction of $\sigma^{\prime}_\mathrm{star}$ for these two effects is not necessary though, for the key to a decisive result of this experiment is a robust measurement of the difference between the velocity dispersion of apparently round and flattened subpopulations.
The fact that these two effects do not depend on the apparent shapes of galaxies indicates a stellar kinematics origin for the dramatic contrast in the line broadening.

This observed apparent shape versus stellar velocity dispersion relation of the representative low-mass galaxies strongly contradicts the oblate disc scenario to explain the abundant apparently flattened galaxies (predominantly at low masses $\mathrm{lg}(\mathcal{M}_{\star}/\mathcal{M}_{\odot})\sim9$) seen in the early Universe.
Rather, the observed relation is consistent with a major population in an intrinsic form more of prolate spindles, for which we can hardly observe their internal stellar motions at their side views.

\autoref{fig:3} highlights the above discriminative conclusion with more data and dynamical models.
The red and blue lines are theoretical predictions for prolate and oblate ellipsoids randomly projected at 2000 inclination angles, given by Jeans Anisotropic Models\cite{2008MNRAS.390...71C} (JAM; using v8.1 \textsc{jampy} software) assuming constant mass-to-light ratio and cylindrically-aligned velocity ellipsoid.
Multi-Gaussian Expansion\cite{1994A&A...285..723E, 2002MNRAS.333..400C} (MGE) models of projected exponential light profiles are used as mass models, with intrinsic minor-to-major axial ratio 0.3 and 0.35 separately for the oblate and prolate ellipsoids.
We measure effective stellar velocity dispersion\cite{2013MNRAS.432.1709C} $\sigma_\mathrm{e}$ from kinematic maps, which is equivalent to the velocity dispersion one would measure out of a spectrum of light integrated within effective radius.
The final red and blue tracks are $\sigma_\mathrm{e}$ at different inclinations normalized to the $\sigma_\mathrm{e}$ at median ellipticity, showing the contrary trends of dispersion with increasing apparent ellipticity.

The blue squares of \autoref{fig:3} are normalized $\sigma_\mathrm{e}$ at given ellipticities for Milky Way analogs\cite{2024ApJ...973L..29W} observed as part of the integral field spectroscopy MaNGA survey, which are known to be oblate discs of typical minor-to-major axial ratios $\sim0.25$, and they follow the JAM predictions in a marked precision.
As for the major high-$z$ galaxy population observed by JWST, the orange squares show their measured velocity dispersion $\sigma^{\prime}_\mathrm{star}$ normalized to the linearly interpolated value at their median ellipticity, separately done for the measurements including (the two marked by central magenta dots) or masking (the other two) the infilling emission lines.
The observations seem to match the predicted properties for prolate spindles in three perspectives: (1) apparent ellipticities are significantly more skewed toward high values compared to oblate discs (see also the Fig. 4 of our previous work\cite{2024ApJ...973L..29W}); (2) projected velocity dispersion decreases with increasing ellipticity; (3) this dispersion drop is drastic over narrow range of ellipticity.

The stellar kinematics of prolate galaxies is dominated by radially biased random motion.
Therefore the experimental result supports the previously established understanding that dynamically cold structures are preferentially formed at lower redshifts, a picture being extensively challenged recently when the JWST revealed flattened systems are interpreted as dynamically cold discs.
Given the assembly history of galaxy populations\cite{2010ApJ...721..193P}, these early low-mass elongated galaxies around and beyond cosmic noon are actively growing to form their tens and hundreds of times more massive descendants we observe in the local Universe, which are mainly rotating disc galaxies including our Milky Way.
Such an evolutionary link marks a dimensional transition of the luminous galactic structures in the Universe over the last ten billion years, from early cosmic times of prevailing prolate galaxies stretching along one dimension, to late times overwhelmed by oblate galaxies which extend toward two dimensions for the most part.

Galaxies form at the potential bottom of dark matter overdensities and grow along with the development of the large-scale cosmic web.
The prevalent elongated galaxies at high redshifts exclusively evidenced by the internal motions of their stars may ultimately ascend to a totally different pattern, as compared to low redshifts, of the dark matter distributions\cite{2015MNRAS.453..408C, 2016MNRAS.458.4477T} at galactic scales and beyond, shedding light on the hierarchical structure formation of Universe over time.
Statistically large samples of galaxies at different redshifts with deep integral field spectroscopy will be required in the future to measure the intrinsic shape of distant individual galaxies and the temporal evolution of its distributions.
With advances in simulations in tandem, then we may be able to reveal the physical origin for this dramatic dimensional transition of galaxy populations, and also to better understand the structural evolution history of our Milky Way which probably went through such transition in the past.


\vspace{+1cm}

\section*{Acknowledgements}

B.W. thanks Michele Cappellari for his kind support and suggestions. Y.P. acknowledges support from the National Natural Science Foundation of China (NSFC) under grant Nos. 12125301, 12192220, and 12192222, and from the New Cornerstone Science Foundation through the XPLORER PRIZE.

\section*{Author contributions}
All contributed significantly to the work presented in this paper.

\section*{Competing interests}
The authors declare no competing interests.





\vspace{+0.5cm}

{\bf Correspondence and requests for materials} should be addressed to B.W. (bitao.wang@hnu.edu.cn).



\appendix

\begin{figure*}
	\includegraphics[width=0.7\textwidth]{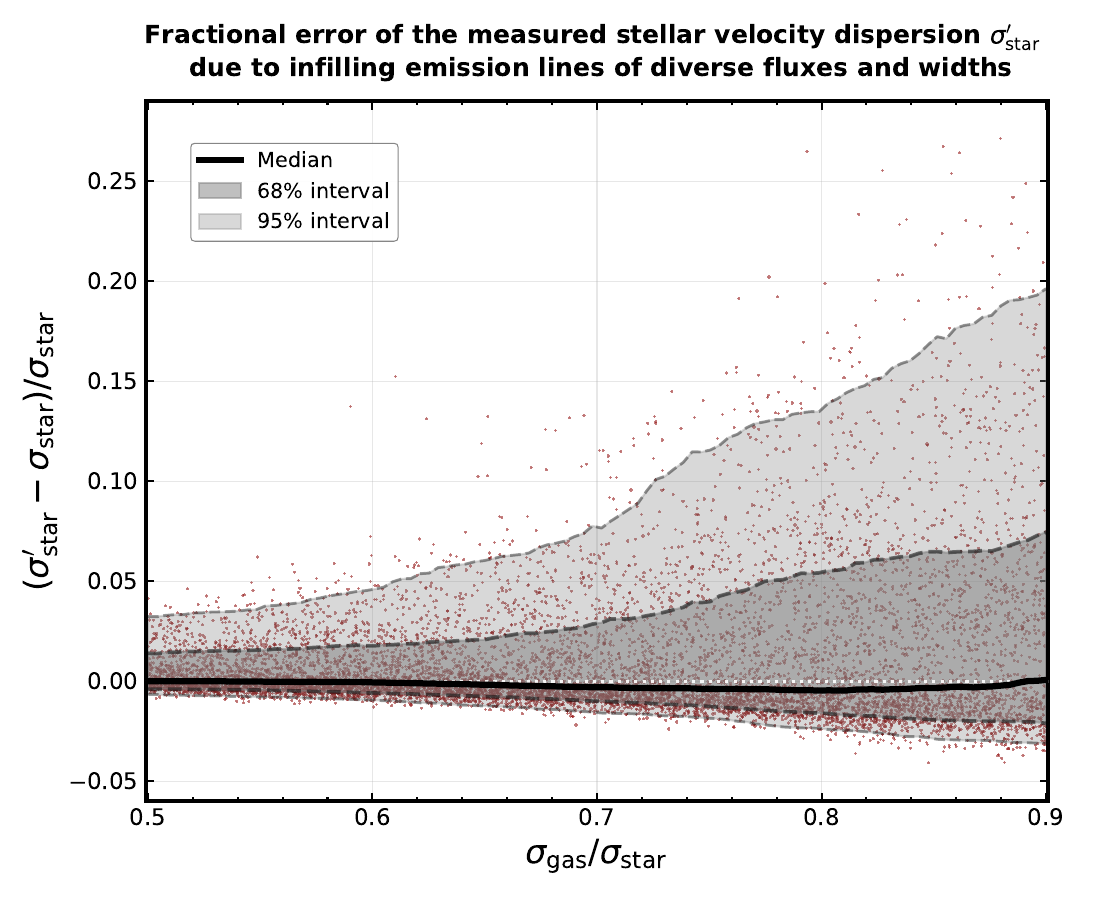}
	\caption{\textbf{Monte Carlo simulation assessment of the errors in the measured stellar velocity dispersion introduced by infilling emission lines.}
  This figure shows the small fractional errors of the measured stellar velocity dispersion $\sigma^{\prime}_\mathrm{star}$ on the y-axis as a function of the intrinsic dispersion ratio between ionized gas and stars on the x-axis for 10k mock spectra, with red dots showing the individual results and the black solid line and the two shaded bands illustrating the median, the 68\% and 95\% interval of the distribution respectively.
	}
	\label{fig:el}
\end{figure*}


\bsp    
\label{lastpage}
\end{document}